\documentclass[%
 reprint,superscriptaddress,
amsmath,amssymb,
prb,
floatfix
]{revtex4-2}

\usepackage{graphicx}
\usepackage{dcolumn}
\usepackage{bm}
\usepackage{xcolor}
\usepackage{soul}
\usepackage{mathtools}

\usepackage{longtable}
\usepackage{rotating,tabularx}

\newcommand\ce[1]{\ensuremath{\mathrm{#1}}}

\begin{document}

\setcitestyle{super}
\title{Thermodynamics of Sodium-Lead Alloys for Negative Electrodes from First-Principles}

\author{Damien K.\ J.\ Lee}
\author{Zeyu Deng}
\affiliation{Department of Materials Science and Engineering, National University of Singapore, 9 Engineering Drive 1, 117575, Singapore}

\author{Gopalakrishnan Sai Gautam}
\affiliation{Department of Materials Engineering, Indian Institute of Science, 560012 Bangalore, India}

\author{Pieremanuele Canepa}
\email{pcanepa@uh.edu}
\affiliation{Department of Electrical and Computer Engineering, Houston, TX 77204, USA}
\affiliation{Texas Center for Superconductivity, University of Houston, Houston, TX 77204, USA}
\affiliation{Department of Materials Science and Engineering, National University of Singapore, 9 Engineering Drive 1, 117575, Singapore}


\begin{abstract}
Metals, such as tin, antimony, and lead (Pb) have garnered renewed attention for their potential use as alloyant-negative electrode materials in sodium (Na)-ion batteries (NIBs). Despite Pb's toxicity and its high molecular weight, lead is one of the most commonly recycled metals, positioning Pb as a promising candidate for a cost-effective, high-capacity anode material.  Understanding the miscibility of Na into Pb is crucial for the development of high-energy density negative electrode materials for NIBs. Using a first-principles multiscale approach, we analyze the thermodynamic properties and estimate the Na-alloying voltage of the Na-Pb system by constructing the compositional phase diagram. In the Pb-Na system, we elucidate the phase boundaries of important phases, such as Pb-rich face-centered cubic and $\beta$-\ce{NaPb_3}, thereby improving our understanding of the phase diagram of the Na-Pb alloy. Due to the strong ordering tendencies of the Na-Pb intermetallics (such as NaPb, \ce{Na_5Pb_2}, and \ce{Na_{15}Pb_4}), we do not observe any solid-solution behavior at intermediate and high Na concentrations.
\end{abstract}

\maketitle

\section{Introduction}
Sodium (Na)-ion batteries (NIBs) present a cost-effective alternative to their lithium (Li)-ion counterparts, primarily owing to the greater natural abundance of Na and its compatibility with aluminum current collectors, as opposed to the more expensive and heavier copper ones.\cite{kim_electrode_2012} The reduced dependence on expensive materials, such as Li or cobalt, positions the NIB technology as particularly well-suited for large-scale energy storage applications, offering an economically viable solution to address the challenges posed by intermittent renewable energy sources.  \cite{vaalma_cost_2018, slater_sodium-ion_2013, palomares_na-ion_2012}.

Unlike in Li-ion electrochemistry, graphite cannot be used as a negative electrode in NIBs since Na does not reversibly intercalate into graphite at low voltages.\cite{ge_electrochemical_1988} Hence, hard carbon-based materials are generally used as the negative electrode for NIBs,\cite{stratford_correlating_2021, stevens_mechanisms_2001} achieving capacities of 300--350~mAh/g.

Sodium-based alloys, specifically with elements of groups 14 and 15, can potentially offer higher capacities than hard carbon materials, in which Na undergoes an alloying reaction with another element to form stable phases. With high theoretical capacities of 847, 660, and 2596~mAh/g, respectively, tin, antimony, and phosphorous have been investigated as negative electrode materials in Na batteries.\cite{hasa_challenges_2021, rudola_commercialisation_2021, stratford_investigating_2017, marbella_sodiation_2018} However, one major drawback of alloy-type electrodes is the large volume change during cycling, causing electrode pulverization, loss of contact with the current collector, and rapid capacity fading.\cite{tian_promises_2021} Further, the active electrode material can fracture and disintegrate, which have been experimentally observed in many \emph{in-situ} studies of lithium-ion battery (LIB) and NIB negative electrodes, such as silicon and tin.\cite{beaulieu_colossal_2001, liu_anisotropic_2011, ying_metallic_2017} 

Despite its inherent toxicity and high molecular weight, lead (Pb) has been used in batteries ever since the invention of the lead-acid battery in 1859. Notably, lead-acid batteries are currently recycled at a rate of 99\%,\cite{may_lead_2018, yanamandra_recycling_2022} charting them as one of the world's most recycled technological devices. Thus, despite the limitations, lead can be explored as a cost-effective alloy-type negative electrode material for large-scale NIB-based energy storage solutions.

So far, there have been limited investigations of using lead as a negative electrode in NIBs. \citeauthor{jow_role_1987} were the first to investigate the electrochemical reaction of Na with Pb,\cite{jow_role_1987} where the authors reported a series of intermetallics that form during the alloying of Na with Pb, following the sequence \ce{Pb \rightarrow NaPb_3 \rightarrow NaPb \rightarrow Na_5Pb_2 \rightarrow Na_{15}Pb_4}.\cite{jow_role_1987} These phases agree well with the experimental phase diagram of the Na--Pb system.\cite{lamprecht_solubility_1968} In its fully sodiated phase, \ce{Na_{15}Pb_4} provides a theoretical capacity of 485~mAh/g.

More recently, \citeauthor{darwiche_reinstating_2016}\cite{darwiche_reinstating_2016} reported noteworthy capacity retention (464 mAh/g) and Coulombic efficiency (99.99\%) over 50 cycles for a 98 wt.\% Pb as an electrode for NIBs. Through \emph{in-situ} X-ray diffraction (XRD), the authors observed successive phase transitions during the sodiation of Pb,\cite{darwiche_reinstating_2016} highlighting the absence of a solid solution behavior. Importantly, the authors reported the formation of an intermediate phase with lower Na content, namely \ce{Na_9Pb_4} (\ce{Na:Pb}~=~2.25), instead of \ce{Na_5Pb_2} (2.50).\cite{darwiche_reinstating_2016}

To date, the extent of spontaneous Na-Pb mixing has not been fully elucidated in the Na-Pb system, particularly in the Pb-rich chemical space.\cite{lamprecht_solubility_1968} In particular, the phase boundaries of the so-called $\beta$-phase, \ce{NaPb_3}, are not resolved over a wide temperature range in the current Na-Pb phase diagram. Thus, we investigate the thermodynamics of the Na-Pb system at varying temperatures, through a multi-scale computational approach. 

We explore the solubility of Na within the Pb face-centered cubic (FCC) structure as a function of temperature, and derive the temperature-dependent voltage profile of the Na-alloying reaction with Pb, through a combination of density functional theory (DFT),\cite{hohenberg_inhomogeneous_1964, kohn_self-consistent_1965} cluster expansion (CE),\cite{sanchez_generalized_1984} and semi-grand canonical Monte Carlo (GCMC) simulations. Importantly, we show evidence of solid solution behavior in the low sodium compositions and also unveil the phase boundaries of phases that occur in the Pb-rich region, which are not well-defined in existing phase diagrams. Our study provides a comprehensive understanding of the thermodynamics of Na alloying with Pb, which unlocks potential strategies for enhancing the reversible capacities and compatibility with electrolytes of alloying-type negative electrode materials in NIBs. 

\section{Results}
\begin{figure*}[!ht]
\includegraphics[width=1.0\textwidth]{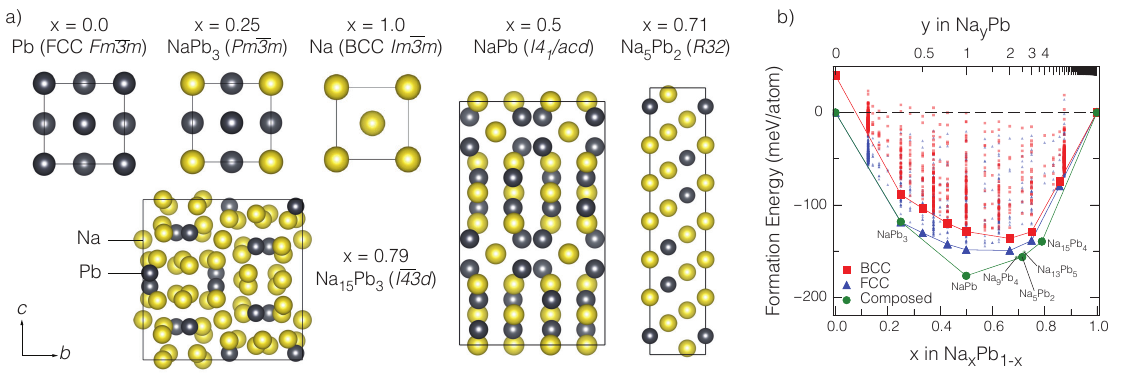}
  \caption{\textbf{a)} Known crystal structures of Na-Pb intermetallics. Na atoms are colored in yellow and Pb atoms in black. x represents Na content in \ce{Na_xPb_{1-x}}. \textbf{b)} Convex hull of the intermetallic phases (green circles), FCC orderings (blue triangles), and BCC orderings (red squares). The references used are the FCC structure for pure Pb and the BCC structure for pure Na.}
  \label{fig:structure}
\end{figure*}

The intermetallic structures of all experimentally known Na-Pb phases are illustrated in Figure \ref{fig:structure}a. Using DFT, the formation energies of all intermetallic Na-Pb orderings are calculated to evaluate their phase stability, which enables the construction of the binary phase diagram at 0 K ---the ``convex hull'' of the Na-Pb system, in Figure \ref{fig:structure}b. The lattice parameters of the optimized structures and the details of our DFT calculations are found in sections S1 and S2 of the Supporting Information (SI), respectively. 

Since Na crystallizes as a body-centered cubic (BCC) structure and Pb in an FCC system, Na-Pb orderings of Pb in BCC Na (red squares in Figure~\ref{fig:structure}b), and Na in FCC Pb (blue triangles), respectively, were considered. In mapping the Na-Pb binary systems with DFT, known Na-Pb orderings (Figure~\ref{fig:structure}a) at intermediate compositions whose structures deviate from either that of BCC Na or FCC Pb were also considered.

The overall convex hull (green circles in Figure~\ref{fig:structure}), minimizing the formation energy of the Na-Pb orderings was constructed with the phases stable at room temperature as reference configurations, namely, FCC for lead and BCC for Na. Thus, the reference configurations for the FCC (BCC) convex hull are pure Pb and pure Na in FCC (BCC) structures. Na-Pb structures forming the convex hull in  Figure~\ref{fig:structure}b are the reflection of stabilizing enthalpic conditions driven by the formation of stable bonds between Na and Pb atoms. 

In Figure~\ref{fig:structure}b, the \ce{Na_9Pb_4} ($P6_3/mmc$) and \ce{Na_{13}Pb_5} ($P6_3/mmc$) phases\cite{ward_three_2015,weston_crystal_1957} appear metastable on the overall convex hull (green line), with relatively small decomposition energies of $\sim$1.51 and $\sim$3.41~meV/atom, respectively. The low decomposition energies indicate that these phases may be entropy stabilized, and hence observed at room temperature.

As illustrated in Figure \ref{fig:structure}b, the BCC convex hull is consistently above both the FCC and the overall convex hulls. The high formation energies of the Na-Pb orderings in the BCC structure indicate the poor thermodynamic stability of this phase, especially for intermediate Na-Pb compositions. This suggests that the BCC-based phases are unlikely to be observed in experiments, apart from high Na-content regions (e.g., x~$>$~0.9 in \ce{Na_xPb_{1-x}}) of the phase diagram.
    
The FCC and the overall convex hull share a common tangent line between x~=~0 and x~=~0.25, which is to be expected since the \ce{NaPb_3} phase ($Pm\bar{3}m$) is an FCC-based ordering, where the corner (face-centered) sites of the cube are occupied by Na (Pb, see Figure~\ref{fig:structure}a). Other Na-Pb orderings in the FCC hull remain above the overall convex hull, indicating the strong thermodynamic stability of the ordered intermetallic phases. In addition, the  FCC polymorph of Na metal is only slightly metastable, by $\sim$0.51~meV/atom, versus its BCC ground-state.

The CE approach,  introduced by \citeauthor{sanchez_generalized_1984}\cite{sanchez_generalized_1984} was used to parameterize the DFT formation energies based on the FCC crystal structure and the associated convex hull. Relying on 975 symmetrically nonequivalent Na-Pb orderings, we constructed a CE to enable the fast evaluation of the formation energy of any arbitrary Na/Pb configuration within the FCC structure.\cite{puchala_thermodynamics_2013, puchala_casm_2023, van_der_ven_first-principles_2018} Details of the CE model are in Section~S3 of the SI. 

The converged CE features 33 unique effective cluster interactions (ECIs) of which are 1 constant term, 1 point term, 6 pairs, 12 triplets, and 13 quadruplets. The transferability of the CE was assessed by the leave-one-out cross-validation (LOOCV) score of the CE model, which was $\sim$4.60~meV/atom. The root-mean-square error (RMSE) of the CE model, $\sim$3.52~meV/atom indicates the quality of the CE fit, with the CE reproducing all ground states predicted by DFT (see Figure~S1 in the Supporting Information). Subsequently, we combined the CE with the semi-GCMC simulations to predict the free energies of an extended range of Na compositions at varying temperatures. Details of the GCMC simulations are found in Section~S4 of the SI.

\begin{figure*}
  \includegraphics[width=1.0\textwidth]{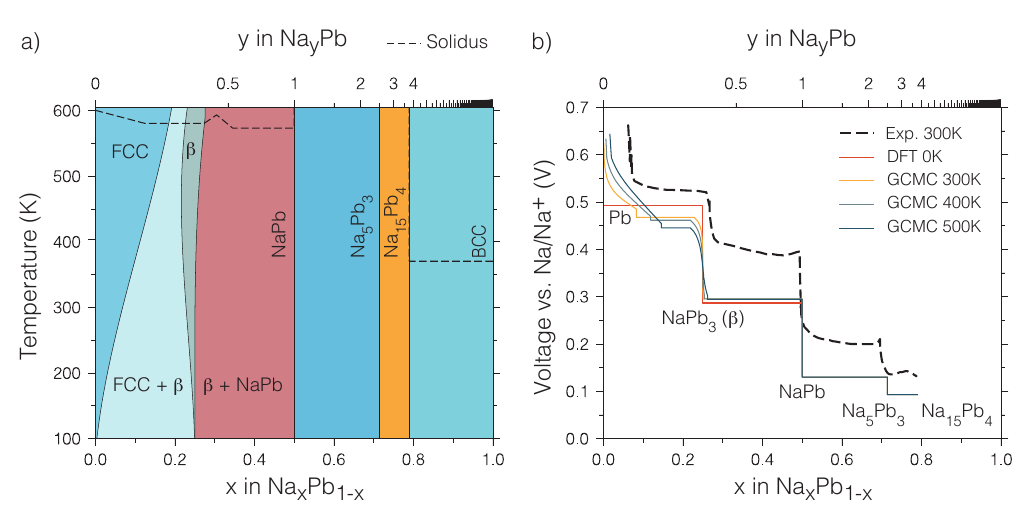}
  \caption{\textbf{a)} Predicted temperature-composition phase diagram of the Na-Pb system. The dashed lines indicate the solidus lines, as taken from the experimental phase diagram of Ref.~\citenum{lamprecht_solubility_1968}. \textbf{b)} Computed voltage (Volts \emph{vs}. Na/Na$^+$) curves as a function of $x$ in \ce{Na_xPb_{1-x}} at different temperatures (T~=~0K, 300K, 400K, and 500K). The experimental voltage of the Na-dealloying from Ref.~\citenum{jow_role_1987} (\textcopyright\/ The Electrochemical Society. Reproduced with permission of IOP Publishing Ltd. All rights reserved) is superimposed on the computed voltage curves as the dashed black line. Since the amount of Pb remains constant during cycling, the number of Na alloyed per Pb atom (i.e., y in \ce{Na_yPb}) is included at the top of both panels for convenience.  }
  \label{fig:phase}

\end{figure*}

The construction of the phase diagram of Figure~\ref{fig:phase}a, and hence the location of the phase boundaries requires the identification of discontinuities in the variation of Na concentration ($x$) versus the sodium chemical potential ($\mu_\mathrm{Na}$). Importantly, we removed the numerical hysteresis arising from GCMC simulations through thermodynamic integration (see section S4 in SI).\cite{deng_phase_2020, puchala_casm_2023} Known ordered Na-Pb intermetallics, such as NaPb, \ce{Na_5Pb_2}, and \ce{Na_{15}Pb_4} of Figure~\ref{fig:structure}b, {were integrated with the GCMC calculations to determine their respective free energies with respect to the FCC phases (see Section S5 of the Supporting Information). This procedure allowed us to obtain a complete phase diagram of Figure~\ref{fig:phase}a.

Remarkably, Figure~\ref{fig:phase}a reproduces with great accuracy the main features of the experimental phase diagram.\cite{lamprecht_solubility_1968} The computed phase diagram of Figure~\ref{fig:phase}a contains two single-phase regions described as: \emph{i}) FCC, Na atoms in the Pb matrix forming a solid solution, \emph{ii}) $\beta$ (\ce{NaPb_3}), a cubic phase consisting of FCC sites with a narrow range of solubility (Figure~\ref{fig:phase}a). The $\beta$ and the FCC Pb solid solution are separated by a miscibility gap. GCMC simulations demonstrated that the miscibility gap closes at temperatures higher than $\sim$650K, which is well beyond the solidus line (dashed black lines) and hence not shown in Figure \ref{fig:phase}a.
    
Other known intermetallics,\cite{pan_perfect_2020} such as NaPb ($I4_1/acd$), \ce{Na_5Pb_2} ($R\bar{3}m$), \ce{Na_{15}Pb_4} ($I\bar{4}3d$) appear as ``line compounds''  on the phase diagram of Figure~\ref{fig:phase}a, since we did not include the effect of off-stoichiometry for these compounds in our calculations. Nevertheless, such compositions also appear as line compounds in the experimental Na-Pb phase diagram.\cite{lamprecht_solubility_1968}

The strong tendency of Na-Pb ordering in the \ce{Na_{15}Pb_4} phase and the significant size difference between Na and Pb atoms suppress the solubility of Pb within BCC Na-metal ($x$~=~1). Therefore, the Na-rich BCC phase appears as a line compound in the phase diagram in the limit $x\rightarrow1$, with no solubility of Pb in Na observed up to 600~K.

Having established the Na-Pb phase diagram, we use this knowledge to derive voltage curves at variable temperatures for Na alloying with Pb, which are useful for guiding the interpretation of electrochemical experiments with the Na-Pb alloy system. Figure \ref{fig:phase}b illustrates the computed alloying voltage profile across the whole Na compositional region at selected temperatures of 0~K, 300~K 400~K, and 500~K. Since the Pb present on the electrode can be assumed to be immobile, and hence the amount of Pb to be constant during battery cycling, we also plot the voltage profile as a function of the number of atoms per Pb atom (i.e., y in \ce{Na_{y}Pb}). 

Note that we obtained the voltage profile at 0~K directly from  DFT calculations of Figure~\ref{fig:structure}b), while voltage profiles at temperatures $>$~0~K  were determined by calculating Gibbs energies through thermodynamic integration (Section~S4 in SI). At 0~K, there are four plateaus, corresponding to the respective two-phase regions: \emph{i}) Pb+\ce{NaPb_3} ($\sim$0.49~V vs.~\ce{Na/Na+}), \emph{ii}) \ce{NaPb_3}+NaPb ($\sim$0.29~V), \emph{iii}) NaPb+\ce{Na_5Pb_2} ($\sim$0.13~V), and \emph{iv}) \ce{Na_5Pb_2}+\ce{Na_{15}Pb_4}.  Expectedly, the plateaus \ce{Na_5Pb_2}+\ce{Na_{15}Pb_4} is located at a very low voltage ($\sim$0.09~V vs.\/{} \ce{Na/Na^+}). 

As the effect of temperature is introduced, subtle variations emerge in the voltage profile compared to the 0~K profile, in particular, in regimes of low Na concentrations (y$<0.5$). For example, at  300~K and 0$<$y$<$0.5, a distinct single-phase region emerges where Na dissolves into the FCC matrix of Pb. The voltage curve for the alloying reaction exhibits a smooth monotonic variation up to the solubility limit of Na. Likewise, the same solid solution behavior is observed corresponding to the single-phase $\beta$ region, occurring within a narrow Na composition range (0.23$<$y$<$0.33). 

With increasing temperature, the width of the two-phase region enclosed between the FCC and the $\beta$ phases in Figure~\ref{fig:phase}a)  diminishes, as indicated by the reduction of the voltage plateau in Figure~\ref{fig:phase}b). However, as elevated temperatures do not expand the width of the line compounds (e.g., \ce{NaPb}, \ce{Na_5Pb_3}, and \ce{Na_{15}Pb_4}, other voltage plateaus in the region 0.35$<$y$<$3.75 remain practically unaffected by temperature.  This is to be expected since the configurational entropy of the ordered intermetallics is 0, resulting in no appreciable temperature effects on the Gibbs energies of formation of these phases.

The experimental voltage profile of the Na-dealloying reaction, reported by \citeauthor{jow_role_1987}\cite{jow_role_1987} is superimposed on the calculated voltage profiles. In general, there is a good (qualitative) agreement between the experimental and theoretical sodiation voltages, with the computed voltage profiles underestimating the experimental data by a systematic offset of $\sim$0.1~V. The magnitude of this offset aligns with the previous report on anode materials for Na batteries.\cite{chevrier_challenges_2011}

\section{Discussion}
In this work, we investigated the thermodynamics governing the alloying of Na into Pb by implementing a multi-scale approach based on first-principles calculations, the cluster expansion formalism, and Monte Carlo simulations. The alloying reaction of Pb with Na holds potential for use in the negative electrode in Na batteries. Unlike intercalation-type electrodes, alloy-type (or conversion-type) electrodes undergo the formation of entirely new phases during reversible sodiation. Understanding the phase changes occurring within the electrode is essential for identifying the limitations imposed by the reaction mechanism, such as practical losses related to initial particle morphology, synthetic approaches, and electrode preparations. \cite{yu_understanding_2018} Here, through the acquired knowledge of the compositional Na-Pb phase diagram, we have identified the thermodynamically stable phases and elucidated the reaction mechanisms occurring during the alloying process manifested in the battery cycling process. Notably, our work sheds light on the phase boundaries of the $\beta$ phase, an important detail that was not identified in previous explorations of the Na-Pb phase diagram.

The computed phase diagram compares favorably with the existing experimentally determined phase diagram. For instance, the Na solubility in the FCC matrices of pure Pb and the $\beta$ phase is observed, while the remaining intermetallic phases \ce{NaPb}, \ce{Na_5Pb_2} and \ce{Na_{15}Pb_4} appear as line compounds.  At low Na concentrations, a solid solution is formed in which Na occupies the FCC sites of the Pb matrix. With increasing temperature, higher configurational entropy results in a lower Gibbs Free energy, stabilizing the mixing process and increasing the solubility limit of Na. The formation of the $\beta$-phase occurs when Na atoms simultaneously occupy the corner sites of the FCC lattice, resulting in an ordered phase with limited solubility.  The miscibility gap between the FCC phase and $\beta$ phase does not close at temperatures below its solidus temperature, which indicates the formation of biphasic regions of differing Na content within the electrode material during sodiation. We speculate that upon sodiation of Pb-FCC, the formation of the $\beta$ phase is likely dominated by vacancy diffusion. Therefore, if the sodiation of the Pb electrode occurs at high rates, thus faster than the time allowed for Na atoms to order, the $\beta$ phase may be suppressed and may not be observed during cycling. Therefore, fast cycling rates may bypass the $\beta$-phase \ce{NaPb_3}, and form a two-phase region between FCC and \ce{NaPb} phases. Suppressing the $\beta$-phase will also increase the solubility limit of Na in the FCC phase.

\begin{figure}[ht!]
  \includegraphics[width=1.0\columnwidth]{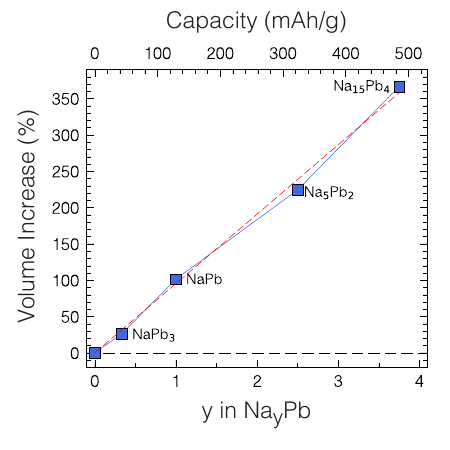}
  \caption{Volume expansion of the alloying reaction (in \% increase) as a function of the amount of Na alloyed per Pb atom. The theoretical capacity of the Pb electrode is indicated. The percent increase is calculated with respect to pure FCC Pb, as indicated by the black dashed line. A linear fit is indicated by the red dashed line, with the equation: Percent Increase~=~95.5~$\times\left(y~\mathrm{in}~\mathrm{Na_yPb}\right)$.}
  \label{fig:volume}
\end{figure}

At Na concentrations $x>$~0.5, it is likely that the metastable \ce{Na_9Pb_4} phase forms instead of the \ce{Na_5Pb_2}, as the voltage step appears closer to $y=2.25$ than $y=2.5$ in \citeauthor{jow_role_1987}'s experimental findings.\cite{jow_role_1987} This finding is corroborated by the in-situ study reported by \citeauthor{darwiche_reinstating_2016}\cite{darwiche_reinstating_2016} There are several reasons why the \ce{Na_9Pb_4} phase is not observed in our results. Firstly, at 0~K, \ce{Na_9Pb_4} does not appear as a ground state but with proximity to the convex hull ($\sim$1.51~meV/atom). With increasing temperature, vibrational entropy might play a role in stabilizing the \ce{Na_9Pb_4} phase. Second, our model does not include vibrational contributions in determining the Gibbs energies. Third, the formation of \ce{Na_9Pb_4} might be more kinetically favorable over \ce{Na_5Pb_2}, which results in its observation during cycling.

At higher Na concentrations, phase separation prevails, with no solid solution behavior due to the formation of ordered intermetallic compounds. Here, the enthalpy of the formation of these compounds dominates, and configurational entropy does not affect the stability of these phases. 
Full sodiation of pure Pb results in the formation of a stable \ce{Na_{15}Pb_4} compound, corresponding to a theoretical capacity of $\sim$485~mAh/g, with 3.75 Na per Pb atom.

Arguably, the primary drawback limiting the use of alloy-type anodes is the pronounced volume change occurring during charge/discharge processes. Upon repeated Na cycling, the substantial volume fluctuation can lead to the pulverization of electrode materials and subsequent capacity fading.\cite{boukamp_allsolid_1981} In Figure \ref{fig:volume}, we quantified the volume expansion of the Pb electrode as a function of the number of Na atoms inserted. The volume increase within the electrode appears nearly proportional to the quantity of Na alloyed with Pb, as evident from Figure~\ref{fig:volume}. A  linear fit of Figure~\ref{fig:volume} demonstrates that the volume expands with a rate of $\sim$95.5\%  per sodium atom alloyed with respect to FCC Pb. The fully sodiated phase of \ce{Na_{15}Pb_4} corresponds to an exceedingly high volume increase of $\sim$366.6\%.

Notwithstanding the large volume variation incurred by Na-Pb alloying reactions, the volume expansion estimated in Figure~\ref{fig:volume} appears lower than that of other commonly-used alloy materials, such as Sn and Sb, which exhibit expansions of 423\% and 390\%, respectively, with respect to the fully sodiated phases (\ce{Na_{15}Sn_4} and \ce{Na_3Sb}).\cite{qiao_advanced_2023} This relatively lower volume change in Pb compared to other alloys suggests its potential utility as a high volumetric energy density material with longer cycle life. Further optimization in particle size, morphology, and blending with other materials, could potentially alleviate the adverse effects of volume changes,\cite{rudola_commercialisation_2021} a strategy that has been successfully employed in silicon nanowires for LIBs,\cite{chan_high-performance_2008, cui_silicon_2021} as well as nanostructures in Sn and Bi-Sb alloys for NIBs.\cite{ni_rooting_2020, ni_durian-inspired_2020}

\section{Conclusions}
The phase behavior of the full compositional space of the Na-Pb binary system was investigated using a multiscale approach. The computed diagram accurately captured the thermodynamically stable intermetallic phases reported in experimental studies. The phase boundaries of the \ce{NaPb_3} phase, are identified, addressing a gap in the known Na-Pb phase diagram. At low Na concentrations, the computed phase diagram highlights the significant solubility of Na within the FCC Pb matrix, especially at higher temperatures, resulting in sloped voltage profiles as Na alloys with FCC Pb. Due to the strong ordering tendencies of the Na-Pb intermetallics (such as NaPb, \ce{Na_5Pb_2}, and \ce{Na_{15}Pb_4}), we do not observe any solid-solution behavior at intermediate and high Na concentrations.

Structural analysis revealed that the Na-Pb alloy expands up to 366.6\% during full sodiation of Pb to \ce{Na_{15}Pb_4}. This expansion is much reduced in magnitude compared to other negative electrode materials, e.g.\ Sn and Sb. Despite its high molecular weight, Pb can deliver a high theoretical capacity of 485~mAh/g, surpassing current commercial hard carbon materials commonly used in Na-ion batteries. Coupled with its high recycling rate, Pb anodes emerge as a promising candidate for a low-cost negative electrode material in bulk or blended with hard carbons. Our work is an important contributor to the design and engineering of alloy-type negative electrodes for NIBs.

\section{Acknowledgement}
P.~C.\ acknowledges funding from the National Research Foundation under its NRF Fellowship NRFF12-2020- 0012. The Welch Foundation is acknowledged for providing P.\ C.\ a Robert A.\ Welch professorship at the Texas Center for Superconductivity.  The computational work was performed on resources of the National Supercomputing Centre, Singapore (https://www.nscc.sg). 



\pagebreak
\widetext
\begin{center}
\textbf{\large Supplemental Materials:  \\ 
Thermodynamics of Sodium-Lead Alloys for Negative Electrodes from First-Principles}
\end{center}
\setcounter{equation}{0}
\setcounter{figure}{0}
\setcounter{table}{0}
\setcounter{page}{1}
\makeatletter
\renewcommand{\theequation}{S\arabic{equation}}
\renewcommand{\thefigure}{S\arabic{figure}}
\renewcommand{\thetable}{S\arabic{table}}

\section{Structure Characteristics of NaPb Intermetallics. }
\begin{table*}[ht!]
    \centering
    
    \begin{tabular*}
    {\textwidth}{@{\extracolsep{\fill}}llccccl@{}}
    \hline
     {\bf Formula} & {\bf Space Group} & {\bf \textit{a}}  & {\bf\textit{b}} & {\bf \textit{c}}  & {\bf V} & {\bf Ref.}\  \\
     \hline
     Pb & $Fm\bar{3}m$ & 5.028 & -- & --  & 31.780
 & \citenum{bouad_neutron_2003}\\
     \ce{NaPb3} & $Pm\bar{3}m$ & 4.941 & -- &-- & 30.154
& \citenum{havinga_oscillatory_1970} \\
     NaPb & $I4_1/acd$ & 10.661 & -- & 18.039 & 32.035
 & \citenum{marsh_crystal_1953}\\
     \ce{Na5Pb2} & $R\bar{3}m$ & 5.546 & -- & 23.224 & 29.462
 & \citenum{weston_crystal_1957}\\
     \ce{Na9Pb4} & $P6_3/mmc$ & 5.513 & -- & 30.009 & 30.378
 & \citenum{ward_three_2015} \\
     \ce{Na13Pb5} & $P6_3/mmc$ & 5.551 & -- &  40.439 & 29.977 & \citenum{weston_crystal_1957}\\
     \ce{Na15Pb4} & $I\bar{4}3d$ & 13.338 & -- & -- & 31.221 & \citenum{lamprecht_solubility_1968}\\
     Na & $Im\bar{3}m$ & 4.192 & -- & -- & 36.837 & \citenum{hull_new_1917}\\

    \hline
    \end{tabular*}
    \caption{Computed lattice parameters (in \AA), space groups, and volume per atom (in \AA$^{3} \cdot \mathrm{atom}^{-1}$) of all Na-Pb intermetallics.}
    \label{tab:my_label}
\end{table*}

\begin{table*}[ht!]
    \centering
    
    \begin{tabular*}
    {\textwidth}{@{\extracolsep{\fill}}llccccl@{}}
    \hline
     {\bf Formula} & {\bf Space Group} & {\bf \textit{a}}  & {\bf\textit{b}} & {\bf \textit{c}}  & {\bf $\Delta$} & {\bf Ref.}\  \\
     \hline
     Pb & $Fm\bar{3}m$ & 4.950 & -- & --  & +1.58
 & \citenum{bouad_neutron_2003}\\
     \ce{NaPb3} & $Pm\bar{3}m$ & 4.888 & -- &-- & +1.08
& \citenum{havinga_oscillatory_1970} \\
     NaPb & $I4_1/acd$ & 10.580 & -- & 17.746 & +1.21
 & \citenum{marsh_crystal_1953}\\
     \ce{Na5Pb2} & $R\bar{3}m$ & 5.540 & -- & 23.150 & --0.43
 & \citenum{weston_crystal_1957}\\
     \ce{Na9Pb4} & $P6_3/mmc$ & 5.470 & -- & 30.410 & --0.27
 & \citenum{ward_three_2015} \\
     \ce{Na13Pb5} & $P6_3/mmc$ & 5.510 & -- &  40.390 & +0.43 & \citenum{weston_crystal_1957}\\
     \ce{Na15Pb4} & $I\bar{4}3d$ & 13.020 & -- & -- & +2.44 & \citenum{lamprecht_solubility_1968}\\
     Na & $Im\bar{3}m$ & 4.300 & -- & -- & --2.51 & \citenum{hull_new_1917}\\

    \hline
    \end{tabular*}
    \caption{Experimental lattice parameters (in \AA), space groups, and deviation from computed lattice parameters ($\Delta$ in \%) of all Na-Pb intermetallics.}
    \label{tab:exp}
\end{table*}

\section{First-principles Calculations}
The Vienna Ab initio Simulation Package (VASP) \cite{kresse_efficiency_1996, kresse_efficient_1996} was used to perform DFT calculations for all structures. Collinear spin-polarized generalized gradient approximation (GGA)-type Perdew-Burke-Ernzerhof (PBE) functional was used to approximate the unknown exchange-correlation contribution in DFT \cite{perdew_generalized_1996}. The PBE functional was selected as it gives a reasonable tradeoff between accuracy and computational costs, and its efficacy for accurate calculations of various Na-based battery materials within our research group.\cite{deng_phase_2020,Wang2022} Notably, PBE enabled us to carry out 1,200 distinct DFT geometry optimizations. The projected augmented wave (PAW) potentials were used for the description of core electrons.\cite{kresse_ultrasoft_1999} The PAW potentials used were Na \texttt{08Apr2002} $3s^13p^0$ and Pb \texttt{08Apr2002} $6s^26p^2$. Valence electrons were represented using plane waves up to an energy cutoff of 520~eV. A $\Gamma$-centered Monkhorst-Pack  $k$-point mesh with a grid density of 5000/(number of atoms) was used, which was determined after appropriate convergence tests. Although these materials are nonmagnetic, all calculations were initialized as ferromagnetic.\cite{monkhorst_special_1976} The total energy of each structure was converged to within 10$^{-5}$~eV/cell, and atomic forces within 10$^{-2}$ eV$\cdot$\AA$^{-1}$.

The enthalpy of Na mixing into Pb was approximated by the formation energy. The formation energy per atom ($\mathrm{E}_f$) of the alloy at any sodium composition ($x$) can be calculated with Eq.~\ref{eqn:formation}.
\begin{equation}
    \mathrm{E}_f = \mathrm{E}_{\ce{Na_xPb_{1-x}}} - [x\mathrm{E}_{\ce{Na}} + (1-x)\mathrm{E}_{\ce{Pb}}] \label{eqn:formation}
\end{equation}
where $\mathrm{E}_{\ce{Na_xPb_{1-x}}}$ is the DFT total energy of the alloy at a sodium atomic composition $x$, $\mathrm{E}_{\ce{Na}}$ and $\mathrm{E}_{\ce{Pb}}$ are the DFT total energies (per atom) of pure Na and pure Pb in their BCC and FCC phases, respectively.

Since Pb is a heavy element, spin-orbit effects might be present in the system. Using the determined ground state structures on the convex hull, we compared the differences between our PBE calculations and the spin-orbit coupling (SOC)-included calculations (Figure \ref{fig:soc}). There is an average of -15.6 meV/atom difference between the SOC calculations and the PBE calculations. This difference is well within the range of DFT formation energy errors,\cite{sun_thermodynamic_2016, wang_framework_2021} and the points on the convex hull were not affected. Therefore, we do not expect the inclusion of SOC effects to significantly change our results, and have elected to perform the rest of our calculations without inclusion of SOC effects.

\begin{figure}
  \includegraphics{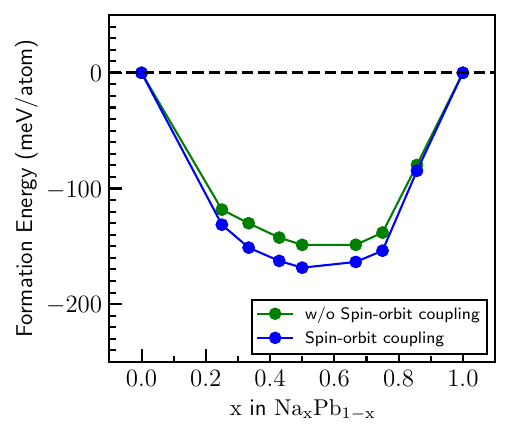}
  \caption{Formation energies of the ground state FCC orderings for the PBE calculations (green) and the PBE-SOC calculations (blue). The average difference between the formation energies obtained with the PBE-SOC calculations and the PBE calculations is --15.6 meV/atom.}
  \label{fig:soc}
\end{figure}

\section{Cluster Expansion Formalism}
In the cluster expansion formalism, the formation energy $\mathrm{E}_f$ of a particular Na-Pb structure is a function of the Na and Pb occupation in a generic FCC lattice, and Eq.~\ref{eqn:ce}.
\begin{equation}
  \mathrm{E}_f(\vec{\sigma}) = V_0 + \sum_i{V_i \sigma_i} + \sum_{i,j}{V_{i,j} \sigma_i \sigma_j} + \sum_{i,j,k}{V_{i,j,k} \sigma_i \sigma_j \sigma_k} 
  \label{eqn:ce}
\end{equation}
where $(V_i, V_{i,j}, \mathrm{and}\ V_{i,j,k})$ are the effective cluster interactions (ECIs) of pair, triplet, and quadruplet clusters, respectively.  $\sigma$ is the occupation variable of each site in the crystal, which is +1 if the site is occupied by Na and --1 if the site is occupied by Pb. Each ECI includes the multiplicity of the cluster and is fit to a set of first-principles training data. All symmetrically distinct pairs, triplets, and quadruplets in the FCC cell within a radius of 12, 8, and 7.5~\AA, respectively, were used to construct the CE model. 

The CE model was fit using the Clusters Approach to Statistical Mechanics (CASM) code.\cite{puchala_thermodynamics_2013, puchala_casm_2023, van_der_ven_first-principles_2018} To fit the CE model, the Na/Pb configurations of different compositions were enumerated up to a maximum supercell size of 8 times the primitive cell. Only structures with a basis deformation of $<0.1$ were included in the CE fit. 0.1 is a typical threshold used to identify structures that match the primitive unit cell. The basis deformation is determined by the mean-square atomic displacement relative to the positions of the atoms in the primitive FCC lattice. During the CE fit, the weights of each structure featured in the CE were optimized using a Bayesian Optimization with Gaussian Processes, which minimized the error between the DFT hull and the predicted CE hull (Figure~\ref{fig:ce}).

The 33 fitted ECIs are displayed in Table \ref{tbl:eci} and plotted in Figure \ref{fig:eci}. A negative ECI indicates energetically favorable interactions between Na and Pb for clusters while a positive ECI indicates unfavourable interactions between atoms of similar species. 

\begin{figure}
  \includegraphics[width=\textwidth]{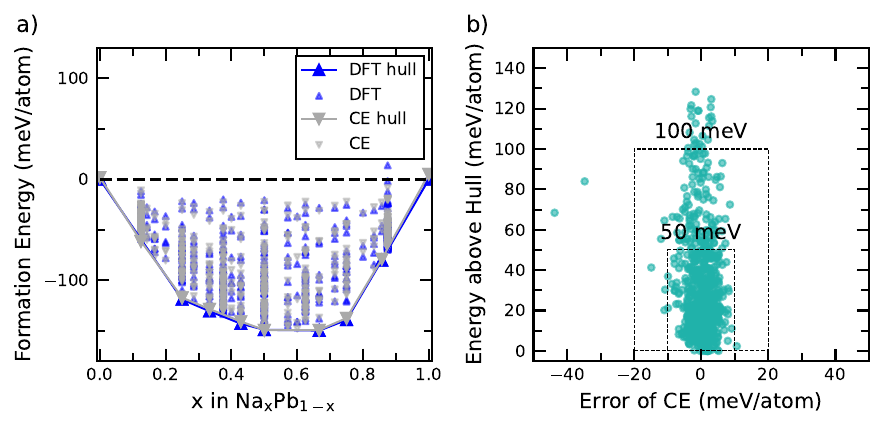}
  \caption{{\bf a)} Formation energies of the FCC orderings (blue) and the formation energies predicted by the cluster expansion model (grey). The references used are the FCC crystal structure for pure Na and Pb. {\bf b)} Error of the CE model with respect to DFT. The dashed lines are the confidence windows ($\pm 10$ meV$\cdot$atom$^{-1}$ and $\pm 20$ meV$\cdot$atom$^{-1}$ for the CE model).}
  \label{fig:ce}
\end{figure}

\begin{sidewaystable}
\centering
    \caption{Effective cluster interactions (ECIs in meV) values of the Point, Pair, Triplet, and Quadraplet terms. Cell $[0, 0, 0]$ is the reference cell. M refers to the multiplicity of each cluster. Maximum (Max.) and Minimum (Min.) cluster lengths are in \AA.}  \label{tbl:eci}
{\footnotesize
\begin{tabular*}{\textwidth}{@{\extracolsep{\fill}}lclcccclr@{}}
    \hline
        \textbf{Index} & \textbf{Cluster Index} & \textbf{Cluster type} & \textbf{Cell} & \textbf{M} & \textbf{Min.} & \textbf{Max.} & \textbf{ECI} & \textbf{ECI/M} \\ \hline
        0 & 0 & Empty & -- & 1 & 0.000 & 0.000 & --113.436 & -113.436 \\ \hline
        1 & 1 & Point & $[0, 0, 0]$ & 1 & 0.000 & 0.000 & --14.220 & --14.220 \\ \hline
        2 & 2 & Pair & $[0, 0, 0] [0, 1, 0]$  & 6 & 3.500 & 3.500 & 119.944 & 19.991 \\ \hline
        3 & 3 & Pair & $[0, 0, 0] [1, -1, -1]$ & 3 & 4.950 & 4.950 & -1.687 & -0.562 \\ \hline
        4 & 4 & Pair & $[0, 0, 0] [1, -2, 1]$ & 12 & 6.062 & 6.062 & --11.092 & -0.924 \\ \hline
        5 & 5 & Pair & $[0, 0, 0] [0, 2, 0]$ & 6 & 7.000 & 7.000 & 6.452 & 1.075 \\ \hline
        6 & 7 & Pair & $[0, 0, 0] [3, -1, -1]$ & 4 & 8.574 & 8.574 & 1.773 & 0.443 \\ \hline
        7 & 10 & Pair & $[0, 0, 0] [0, 3, 0]$ & 6 & 10.501 & 10.501 & --2.864 & --0.477 \\ \hline
        8 & 14 & Triplet & $[0, 0, 0] [0, 0, 1] [0, 1, 0]$ & 8 & 3.500 & 3.500 & 9.574 & 1.197 \\ \hline
        9 & 15 & Triplet & $[0, 0, 0] [0, 0, 1] [1, -1, 1]$ & 12 & 3.500 & 4.950 & 10.437 & 0.870 \\ \hline
        10 & 16 & Triplet & $[0, 0, 0] [0, 0, 1] [1, 0, 1]$ & 24 & 3.500 & 6.062 & --12.251 & --0.510 \\ \hline
        11 & 17 & Triplet & $[0, 0, 0] [0, 1, 0] [1, -1, 1]$ & 24 & 3.500 & 6.062 & -18.552 & -0.773 \\ \hline
        12 & 18 & Triplet & $[0, 0, 0] [0, 1, 1] [1, 1, 0]$ & 24 & 3.500 & 6.062 & 4.501 & 0.188 \\ \hline
        13 & 20 & Triplet & $[0, 0, 0] [0, 1, -2] [0, 2, -1]$ & 8 & 6.062 & 6.062 & --2.336 & --0.292 \\ \hline
        14 & 23 & Triplet & $[0, 0, 0] [0, 0, 1] [0, 2, -1]$ & 48 & 3.500 & 7.000 & 24.069 & 0.501 \\ \hline
        15 & 24 & Triplet & $[0, 0, 0] [0, 1, 1] [2, 0, 0]$ & 12 & 6.062 & 7.000 & --6.514 & --0.543 \\ \hline
        16 & 25 & Triplet & $[0, 0, 0] [0, 0, 2] [0, 2, 0]$ & 8 & 7.000 & 7.000 & --1.651 & --0.206 \\ \hline
        17 & 28 & Triplet & $[0, 0, 0] [0, 1, -2] [1, -1, -2]$ & 48 & 6.062 & 7.827 & 5.966 & 0.124 \\ \hline
        18 & 29 & Triplet & $[0, 0, 0] [0, 1, -1] [2, 0, 0]$ & 24 & 3.500 & 7.827 & 7.520 & 0.313 \\ \hline
        19 & 31 & Triplet & $[0, 0, 0] [1, -2, 2] [2, -2, 1]$ & 24 & 3.500 & 7.827 & --3.592 & --0.150 \\ \hline
        20 & 34 & Quadruplet & $[0, 0, 0] [0, 0, 1] [0, 1, 0] [1, 0, 0]$ & 2 & 3.500 & 3.500 & 1.178 & 0.589 \\ \hline
        21 & 35 & Quadruplet & $[0, 0, 0] [0, 1, -1] [0, 1, 0] [1, 0, 0]$ & 12 & 3.500 & 4.950 & 6.423 & 0.535 \\ \hline
        22 & 37 & Quadruplet & $[0, 0, 0] [0, 0, 1] [1, 0, 0] [1, 0, 1]$ & 12 & 3.500 & 6.062 & 6.600 & 0.550 \\ \hline
        23 & 40 & Quadruplet & $[0, 0, 0] [0, 0, 1] [0, 1, -1] [1, -1, 0]$ & 24 & 3.500 & 6.062 & 4.028 & 0.168 \\ \hline
        24 & 42 & Quadruplet & $[0, 0, 0] [0, 1, 0] [1, -1, 1] [1, 0, 1]$ & 6 & 3.500 & 6.062 & 2.417 & 0.403 \\ \hline
        25 & 44 & Quadruplet & $[0, 0, 0] [0, 1, -2] [0, 1, -1] [0, 2, -1]$ & 8 & 3.500 & 6.062 & --3.835 & --0.479 \\ \hline
        26 & 45 & Quadruplet & $[0, 0, 0] [0, 1, -1] [0, 1, 0] [2, 0, -1]$ & 8 & 3.500 & 6.062 & 3.397 & 0.425 \\ \hline
        27 & 47 & Quadruplet & $[0, 0, 0] [0, 1, 0] [1, 1, -2] [2, 0, -1]$ & 12 & 3.500 & 6.062 & --4.748 & --0.396 \\ \hline
        28 & 55 & Quadruplet & $[0, 0, 0] [0, 0, 1] [0, 2, -1] [1, 1, -1]$ & 24 & 3.500 & 7.000 & --4.435 & --0.185 \\ \hline
        29 & 61 & Quadruplet & $[0, 0, 0] [0, 0, 1] [0, 1, -1] [2, 0, -1]$ & 48 & 3.500 & 7.000 & 6.889 & 0.144 \\ \hline
        30 & 63 & Quadruplet & $[0, 0, 0] [1, -1, 1] [1, 1, -1] [2, 0, 0]$ & 3 & 4.950 & 7.000 & 2.593 & 0.864 \\ \hline
        31 & 64 & Quadruplet & $[0, 0, 0] [1, -2, 1] [1, 0, -1] [2, -2, 0]$ & 12 & 3.500 & 7.000 & --3.203 & --0.267 \\ \hline
        32 & 68 & Quadruplet & $[0, 0, 0] [0, 0, 2] [0, 2, 0] [2, 0, 0]$ & 2 & 7.000 & 7.000 & --0.720 & --0.360 \\ \hline
 \end{tabular*}
}
\end{sidewaystable}

\begin{figure}
  \includegraphics[width=\textwidth]{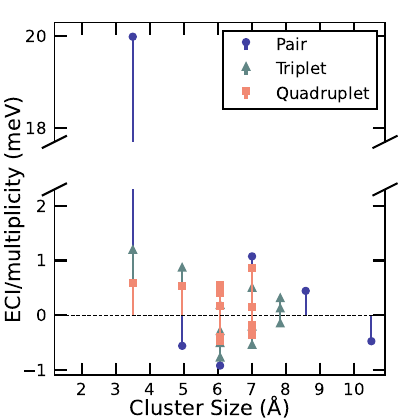}
  \caption{33 effective cluster interactions in meV/multiplicity vs.\ their length (in \AA, $x$-axis). Negative (positive) ECIs in pairs are favourable (unfavourable) interactions between Na-Pb (Na-Na, Pb-Pb).}
  \label{fig:eci}
\end{figure}

\section{Grand-canonical Monte Carlo Simulations}
In a semi-grand canonical ensemble, the composition and energy of the system with a fixed number of sites were sampled while the temperature ($T$) and the Na chemical potential $\mu_\mathrm{Na}$ were varied. The chemical potentials were referenced to the pure FCC Pb and Na phases, for which $\mu_{Na}$ = $\mu_{Pb}$ = 0. Grand canonical Monte Carlo simulations were performed using the CASM package. A $16\times16\times16$ (4,096 atoms) supercell was used for these simulations. GCMC scans were performed in the chemical potential ($\mu$) and temperature (T) space. The scan started at T = 5K and up to 705K with a step $\Delta$T = 5K at $\mu$ = –-0.6, --0.3, --0.1, 0.1, 0.4, and 0.6~eV/atom. These values of $\mu$ correspond to the initial six ground-state structures on the FCC convex hull ($x=0, 0.25, 0.5, 0.67, 0.75, 1$). At every T, $\mu$ was scanned in both forward ($\mu$~=~--0.6 to 0.6~eV/atom) and backward ($\mu$~=~0.6 to --0.6~eV/atom) directions with a step size of $\Delta\mu$ = 0.01 eV/atom  In general, $\mu$ was scanned across 5 concentration ranges ($-0.6<\mu<-0.3, -0.3<\mu<-0.1, -0.1<\mu<0.1, 0.1<\mu<0.4, 0.6<\mu<0.6$).

\section{Thermodynamic Integration}
The grand canonical potential energy ($\Phi$) is defined in Eq.~\ref{eqn:phi}:
\begin{equation}
    \Phi = E-TS-\mu x
    \label{eqn:phi}
\end{equation}
where $E$ is the total energy predicted by the CE model, $S$ is the configurational entropy, and $\mu$ is the parametric chemical potential set in GCMC scans. From GCMC scans at fixed $\mu$ and variable $T$, $\Phi$ is calculated using the thermodynamic integration of Eq.~\ref{eqn:var_T}.
\begin{equation}
\beta\Phi(T,\mu)=\beta_0\Phi(T_0,\mu)+\int_{\beta_0}^{\beta} [E-\mu x] \, d\beta
    \label{eqn:var_T}
\end{equation}
where $\beta$ is the reciprocal temperature, $1/k_B T$, $k_B$ is the Boltzmann constant, and $\beta_0=1/k_B T_0$. $T_0$ is the reference temperature and is chosen to be 5K, in which entropic effects are negligible and hence $\Phi(T_0,\mu)=E-\mu x$.
For GCMC scans at fixed $T$ and variable $\mu$, $\Phi$ is calculated using thermodynamic integration in Eq.~\ref{eqn:var_u}:
\begin{equation}
    \Phi(T,\mu)=\phi(T,\mu_0)-\int_{\mu_0}^{\mu} x \, d\mu
    \label{eqn:var_u}
\end{equation}
 Thermodynamic integration was carried out numerically with the composite trapezoidal rule.

For ordered intermetallic phases, the grand-canonical free energy can be obtained directly from Eq.~\ref{eqn:phi}.
\begin{equation}
    \label{eqn:phi}
    \Phi=E-\mu x
\end{equation}
since the configurational entropy ($S$) of an ordered phase is ~0. 

The grand canonical energies for each phase can be plotted on a single plot to determine the phase boundaries. The phase boundaries are identified from the intersections of $\Phi$ for each phase in the $\mu$ space and from the discontinuities in $x$ vs.\ $\mu$ plots.

\section{Voltage Calculations}
The (de)alloying reaction of Na into(from) Pb metal can occur through the reversible reaction Eq.~\ref{eqn:chem_reaction}:
\begin{equation}
    \mathrm{Na}_x\mathrm{Pb} + (y-x)\mathrm{Na} \xleftrightarrow{-\Delta G} \mathrm{Na}_y\mathrm{Pb}
    \label{eqn:chem_reaction}
\end{equation}
where $x$ and $y$ represent the initial and final Na content per Pb atom, and $-\Delta G$ is the change in Gibbs free energy of the reaction.

The voltage of alloying a particular composition of Na in Pb metal is calculated using Eq.~\ref{eqn:voltage}:
\begin{equation}
\label{eqn:voltage}
    V = -\frac{\Delta G}{nF}
      = -\frac{G_{f,\mathrm{Na_yPb}} - G_{f,\mathrm{Na_xPb}}}{y-x}
\end{equation}
where $G_f$ is the Gibbs free energy (in eV) to form a Na-Pb alloy at a particular composition, and is calculated after thermodynamic integration of the grand-canonical potential at temperature $T$:
\begin{equation}
    G=E-TS
     =\Phi+\mu x
\end{equation}
\bibliography{references}
\end{document}